# Annihilation and control of chiral domain walls with magnetic fields

Sunil K. Karna[1,2,*], Madalynn Marshall[3], Weiwei Xie[3], Lisa DeBeer-Schmitt[4], David P. Young[1], Ilya Vekhter[1], William A. Shelton[5], Andras Kovács[6], Michalis Charilaou[7], John F. DiTusa[1,8,*]

[1]Department of Physics & Astronomy, Louisiana State University, Baton Rouge, LA 70803, USA
[2]Department of Physics & Center for Materials Research, Norfolk State University, Norfolk, VA 23504, USA
[3]Department of Chemistry, Louisiana State University, Baton Rouge, LA 70803, USA
[4]Neutron Scattering Division, Oak Ridge National Laboratory, Oak Ridge, Tennessee 37831, USA
[5]Cain Department of Chemical Engineering, Louisiana State University, Baton Rouge, LA 70803, USA
[6]Ernst Ruska-Centre for Microscopy and Spectroscopy with Electrons, Peter Grünberg Institute, Forschungszentrum Jülich, 52425 Jülich, Germany
[7]Department of Physics, University of Louisiana at Lafayette, Lafayette, LA 70504, USA
[8]Department of Physics, Indiana University-Purdue University Indianapolis, Indianapolis, IN 46202, USA

**The control of domain walls is central to nearly all magnetic technologies, particularly for information storage and spintronics. Creative attempts to increase storage density need to overcome volatility due to thermal fluctuations of nanoscopic domains and heating limitations. Topological defects, such as solitons, skyrmions and merons, may be much less susceptible to fluctuations, owing to topological constraints, while also being controllable with low current densities. Here we present the first evidence for soliton/soliton and soliton/antisoliton domain walls in the hexagonal chiral magnet $Mn_{1/3}NbS_2$ that respond asymmetrically to magnetic fields and exhibit pair-annihilation. This is important because it suggests the possibility of controlling the occurrence of soliton pairs and the use of small fields or small currents to control nanoscopic magnetic domains. Specifically, our data suggest that either soliton/soliton or soliton/antisoliton pairs can be stabilized by tuning the balance between intrinsic exchange interactions and long-range magnetostatics in restricted geometries.**

*Keywords*: Soliton pair dynamics, Nanoscopic chiral domain walls, Dzyaloshinskii-Moriya interaction, Chiral magnets, and Shape anisotropy.

A dramatic increase in investigations of magnetism in materials having a chiral crystal structure followed the discovery of regular arrays of skyrmions, whirls of the local magnetization all with the same chirality (handedness) arranged in a lattice, in MnSi [1-7]. These suggested a new route towards overcoming domain wall volatility, the random switching of small magnetic domains due to thermal fluctuations [8-15], which may be enhanced in low-dimensional structures with magnetic anisotropy and where the interplay of topology and thermal nucleation has long been realized [16]. Further exploration of MnSi made clear that the underlying crystal symmetry plays a dominant role in determining the magnetic states that emerge in this and similar cubic crystals with the *B20* symmetry [1,3,7]. The small crystalline anisotropy and cubic symmetry of the *B20*'s are essential to the formation of helical domains in the ground state, as well as the conical and skyrmion lattice states that appear with small rotational-symmetry-breaking magnetic fields. In contrast, the reduced symmetry and related crystalline anisotropy found in hexagonal chiral magnets results in a very different set of magnetic states [7,17]. Here, the helical pitch in the magnetically ordered state is confined to the crystallographic *c*-axis, even when exposed to a magnetic field, $H$. Thus, for $H$ lying perpendicular to the *c*-axis, skyrmion lattices are not found. Instead, experiments indicate a distorted helical structure allowing the possibility of the formation of chiral magnetic soliton domain walls (DW) [4,18,19]. How this picture changes with variations in the physical parameters that control the size and character of DW has yet to be fully explored [20].

One route towards producing hexagonal chiral magnets has been to intercalate transition metal elements between the hexagonal layers of van der Waals compounds [4,12-23]. Here, we intercalate the 3d transition metal Mn into $NbS_2$ (Fig. 1a), forming a magnet with a crystal structure that lacks both inversion and mirror symmetries. The magnetic properties are explored through magnetometry, small angle neutron scattering (SANS), and Fresnel imaging in Lorentz transmission electron microscopy (LTEM). We compare these data with predictions of models and micromagnetic simulations that corroborate the discovery of a linear soliton lattice and the observation of soliton-antisoliton annihilation by an external magnetic field, confirming theoretical predictions [5]. Here, a soliton with opposite handed modulation is referred to as an antisoliton to distinguish from the homochiral case.

While our previous investigations of $Mn_{1/3}NbS_2$ revealed moments lying along the $NbS_2$ planes forming a nearly ferromagnetic state below $T_c$=45 K [21,23], Fresnel defocused [24] images taken on thin lamella (Fig. 1) display ferromagnetic domains of 100's of nanometers in size with chiral (Bloch) DW. The DW propagate along the crystallographic *c*-axis with a rotation in the NbS$_2$ plane (Fig. 1(b, c)). These Fresnel micrographs of Mn$_{1/3}$NbS$_2$ differ significantly from what was found in isostructural Cr$_{1/3}$NbS$_2$ where a simple helimagnetic state with equally spaced bright and dark stripes was observed in LTEM for a thin lamella [4]. Furthermore, Fig. 1(b, d and f) demonstrate a dramatic change with a reduction of the thickness, *t*, of the lamella. For the relatively thick specimens, $t$=230 nm (Fig. 1(b, c)), a nearly periodic sequence of bright lines interspaced between darker regions at distances of ~250 nm along the *c*-axis appeared when images were taken at 12 K. These micrographs change significantly for thinner lamella, $t$~130 nm Fig 1(d) and $t$~160 nm Fig. 1(f), where alternating bright and dark stripes are separated by grey regions of roughly 1 μm along the *c*-axis. Again, strict periodicity is not observed. For the three samples that we have measured, warming above 25 K causes a loss of contrast, and subsequent cooling results in a similar pattern of stripes, albeit at a different location within the field of view [See Video S1 in Supplemental Information (SI)]. This suggests a magnetic origin for the contrast, a conclusion strengthened by the sensitivity of the contrast pattern to small field as demonstrated in frames (g)-(k). Keeping in mind that LTEM is only sensitive to magnetic moments lying in the plane normal to the electron beam (i.e. the lamella plane), the contrast pattern in Fig. 1(b) implies a rotation of the magnetization within the hexagonal *ab*–plane of the crystal as highlighted by the sharp bright stripes. This image is consistent with a distorted helical magnetic structure where magnetic moments tend to lie in the plane of the lamella modifying this easy-plane system toward an effective easy-axis one. The appearance of alternating dark and bright stripes in the thinner samples (Fig. 1(d, f)) separated by larger regions of slowly varying or nearly constant contrast is substantially different from that seen in the thicker sample (Fig. 1(b)) or in Cr$_{1/3}$NbS$_2$ [4] requiring a different interpretation.

Perhaps more intriguing is the response of the contrast pattern to small *H* oriented parallel to the electron beam, Fig. 1(g-k) (and at several other fields in Figs. S2-S4)). For fields of one sign (defined positive here), dark stripes are seen to translate rightward and bright stripes leftward until they approach each other above 30 mT forming dark/bright pairs. For larger *H* (Fig. 1(i)) they begin to annihilate each other with vestiges of the pairs apparent at the edge of the sample so that the contrast persists at the upper edge. The contrast lines that persist merge at a distance of a few hundred nanometers from the edge where the contrast is lost. Significant hysteresis is apparent as *H* decreases (Fig. S2(j)) until the direction switches (negative *H*), causing the reappearance of alternating dark and bright stripes. These stripes move in an opposite direction as the *H* is increased in the negative sense, forming tight bright/dark pairs (Fig. S2(k) and Fig. S3). This unusual asymmetry in the motion of chiral DW is not yet understood. However, it is likely a consequence of the Dzyaloshinskii-Moriya interaction (DMI) on DW and interactions between them similar to that observed in ferromagnetic films with perpendicular magnetic anisotropy [25]. Alternatively, it may be a consequence of the variation of sample thickness along the lamella.

Insight into these results are made by considering a model where the total energy density contains contributions from the exchange stiffness, $A$, easy-plane anisotropy $K$, DMI, $D$, coupling to the external magnetic field, $\mathbf{H}$, and the dipole-dipole interactions via a local demagnetizing field $\mathbf{H_{dm}}$:

$$\mathcal{E} = A(\nabla \mathbf{m})^2 + K(m_c)^2 + D\mathbf{m} \cdot (\nabla \times \mathbf{m}) - \mu_0 M_s \mathbf{H} \cdot \mathbf{m} - \frac{1}{2}\mu_0 M_s \mathbf{H_{dm}} \cdot \mathbf{m}. \qquad (1)$$

Here, $\mathbf{m} = \mathbf{M}/M_s$ is the magnetization unit vector with $M_s$ the saturation magnetization. We use this model both to perform the full micromagnetic simulations (see Methods) and to understand the main features of the experimentally observed structure using a simplified continuum description. In the latter approach, we take $K$ to be large enough, so that the spins are always in the easy plane, while the DM modulation vector is along the hard axis with $\mathbf{m}(z) = (\cos\phi(z), \sin\phi(z), 0)$. To make analytic progress we replace the demagnetization term with the effective in-plane anisotropy, $\widetilde{K}$, that increases with decreasing thickness of the sample, and favors spins in the plane of the lamellae, i.e. with the term $\widetilde{K}\sin^2\phi$. This approach neglects edge effects that are captured by the full simulations, but is adequate for classifying the phases of the model.

Under these assumptions, for field in the easy plane but normal to the lamellae, the phase $\phi(z)$ satisfies the double sine-Gordon (dSG) equation, $2A\phi_{zz} - \widetilde{K}\sin 2\phi + H\cos\phi = 0$. The energy of the solutions is modified by the DMI, which distinguishes this problem from other physical contexts where the dSG appears [26-27]. For $\widetilde{K} = H = 0$ we recover the well-known helical state, $\phi(z) = q_0 z$ with $q_0 \sim D/2A$. A much longer pitch of the helix in $Mn_{1/3}NbS_2$ compared to $Cr_{1/3}NbS_2$ (~250nm [21] vs 48nm [28]) indicates a smaller DMI strength, and therefore greater role of the dipolar-driven anisotropy. For $\widetilde{K} \neq 0, H = 0$ the spins prefer to be in the plane of the lamellae, $\phi = 0, \pi$, and these two classical configurations are connected by Bloch DW, which are the solutions of the sine-Gordon equations for the phase $\phi(z)$. The DMI interaction lowers (raises) the energy of these DW to be $E_{sK\pm} \sim \sqrt{2A\widetilde{K}} \pm D$ depending on the chirality (winding number, $w = \frac{1}{2\pi}\int_{-\infty}^{+\infty} \phi_z dz = \pm 1$). Therefore, for anisotropies $0 < \widetilde{K} \leq D^2/2A$ the ground state of the system is the lattice of chiral Bloch $\pi$-DW. This agrees with the results of simulations presented in Figs. 2(a,b) and likely corresponds to the LTEM data in Fig. 1(b,c). The origin of this state is similar to that appearing for $\widetilde{K} = 0$ under a finite field, where the energy of $2\pi$ solitons (vs $\pi$ DW) is $E_{sh\pm} \sim \sqrt{2AH} \pm D$, so that a chiral soliton lattice is stabilized for $H \leq H_c \sim D^2/2A$ [29]. This lattice has been observed in $Cr_{1/3}NbS_2$ [30].

For higher anisotropy (thinner samples), the DW are either thermally generated or pinned by the boundaries, and the DMI-induced difference in the energies of DW of different winding is small compared to the domain-wall energy. Then, the field at the lateral edges of the lamella is essential, and the description of dipolar interactions as leading to an effective uniaxial anisotropy is insufficient. In simulations, at *H=0* we find wide regions of spins tilted slightly away from the plane, separated by the DW with spins normal to the lamella in the opposite direction, see Figs. 2(e,g,h). The total winding number is determined by the boundary conditions, and for topologically trivial boundaries, DW appear mostly in pairs adding up to $w = 0$. A sequence of red stripes (magnetic moments pointing up at each of the DW) in Fig. 2(e) indicates switching chirality between sequential domains, see Fig. 2(g,h), an absence of net winding, and hence, non-topological nature of the magnetic order. This should be contrasted with the quasiperiodic red/blue pattern in Fig. 2(a) characteristic of the chiral state.

When the field $H < H_c = 2\widetilde{K}$ is applied normal to the lamellae the DW-like kinks connect classical spin configurations tilted from plane by the angle $\phi_0 = \sin^{-1} H/2\widetilde{K}$. Small (large) kinks have phase varying in the regions $(\phi_0, \pi - \phi_0)$ and $(-\pi - \phi_0, \phi_0)$ respectively [31]. Similar phenomena (without DMI) have been predicted and analyzed in the B-phase of $^3$He [32, 33]. Above the critical field spins are polarized, small kinks vanish, and the energy of the large kinks continuously transforms into that of the $2\pi$ soliton known from $\widetilde{K} = 0$. This is confirmed by the simulations for a thick sample, Fig. 2(c,d), (as well as video SM5) showing the chiral pattern similar to that observed in Ref. [4].

Small kinks have spins nearly aligned with the field, hence they have lower energy and higher density at moderate fields, as is clear from simulations, Fig. 2(f): the light red regions (moments tilted towards the field) are mostly separated by bright red regions (moments along the field). Importantly, since the dSG equation is not exactly integrable, these kinks interact as they are not exact eigenstates of the system at any field [31,34,35]. Studies in the absence of the DMI demonstrated trapping of kink-antikink pairs into long-lived quasi-bound states [34,35] equivalent to non-topological magnetic bions [36]. Experimental observation of pairs of bright and dark lines in LTEM patterns under a magnetic field, Figs. 1(h,k), suggests that the DMI interaction may help stabilize these pairs. Vanishing of the signal at higher fields, once the lines approach each other, indicates that these are objects with opposite winding numbers, so that the global state is non-topological. We note that the general features of the structure and the field-dependence of the observed states are reminiscent of those predicted for thin ferromagnetic films in Ref. [37], but the role of the DM interaction, not accounted for in that analysis, needs to be fully elucidated theoretically.

To place these images and calculations in context and to better establish the magnetic state of the system from which the domain structures imaged in Fig. 1 derive, we have measured the magnetic properties of bulk single crystals adding more understanding to previous results [21,23,38]. These established a magnetic phase transition to a nearly ferromagnetic state below $T_c$ via *dc* magnetization measurements and neutron diffraction. Our new measurements suggest a phase diagram shown in Fig. 3(a), where we highlight a distinct change in behavior below ~25 K (phase I). The response of this system to magnetic fields as observed in the magnetization, $M(H)$, and the *ac* susceptibility, reveals changes not commonly observed in simple magnets. For example, $M(H)$ with $H$ oriented perpendicular to the *c*-axis is displayed in Fig. 3(b) where a hysteresis is apparent for $T < 25$ K only for non-zero $H$, illustrated in Fig. 3(a) by dotted lines. The maximum temperature where this hysteresis is found, $T$~25 K, corresponds with distinct changes in the $T$ and $H$ dependent *ac* susceptibility shown in Figs. 3(c-f). Most dramatic is the reduction in the imaginary part of the *ac* susceptibility, $\chi''$, at all $H$ for $T < 25$ K (Fig. 3(d,f)). Thus, phase I is characterized by the hysteresis in $H$ and the small $\chi''$, corresponding well with the $T$ and $H$ region where lines of contrast were observed in the Fresnel images. The implication is that there is a distinct change in the magnetic domain structure and dynamics at the boundary of phase I with phases II and III since the range of $H$ spanning the purported phase I typically corresponds to mesoscopic-sized features. This conclusion is supported by the variation of the frequency dependence of $\chi'$ and $\chi''$ as displayed in Fig. S6.

The other regions of the phase diagram are categorized by the response observed in $\chi'$ and $\chi''$ including for $T > T_c$, where a small $\chi'$, and $\chi''$ are consistent with a paramagnetic state. For temperatures between 24 and 45 K and $H < 40$ mT (phase II in Fig. 3(a)), the response is characterized by a highly $H$-dependent $\chi'$ and a large $\chi''$ that is maximum near fields when $M(H)$ approaches saturation. Finally, for $H > 40$ mT the system is nearly saturated. However, since a peak in $\chi'$ persists at the transition between phase III and the PM state ($T_1$) at $H$ well above the apparent saturation, and $\chi''$ continues to evolve at these higher $H$, we hesitate to refer to this region as fully field polarized.

Further insight to the magnetic structure was accomplished through small angle neutron scattering (SANS) measurements (Fig. 4). The geometry of the measurement has the crystallographic $c$-axis nearly horizontal in the plane of the detector, while the neutron beam lies along the $ab$-plane. For $T < T_C$, Fig. 4 (a-c), a streak of scattering along the $c$-axis is apparent and increases in intensity with temperature near scattering vector, $Q=0$, particularly for $T > 32$ K. These data are presented in graphical form in Fig. 4(e), where the intensity after integration between azimuthal angles, $\chi_{az}$, lying within the white, wedged shaped, regions in Fig. 4 (a-c) is plotted as a function of $Q$, and in Fig. S9. The variations we observe with cooling are likely related to the evolution of the ac susceptibility that motivated the phase diagram of Fig. 3(a). This scattering streak signals a disordered magnetic structure consisting of either ferromagnetic domains or a non-sinusoidal helical magnetic structure. Whether the disorder is intrinsic to $Mn_{1/3}NbS_2$, or is a result of Mn site defects [21] or the possible presence of stacking faults along the $c$-axis evident in x-ray diffraction of larger crystals, is not yet known. However, a simple disordered ferromagnetic state is not likely since the width of magnetic scattering along the $c$-axis would resemble that found in the nuclear Bragg scattering. The crystallographic disorder apparent in our previous neutron diffraction measurements [21] and our single and powder crystal x-ray characterization is not compatible with the mosaicity required by the SANS data for a disordered ferromagnetic state. In addition, high resolution electron micrographs of our LTEM specimens display minimal disorder, with no indication of stacking faults on the scale of the images, and no indication of an incommensurate order (Fig. S8). Instead, the SANS data indicate a non-periodic stripe phase, a conclusion driven by the Fresnel images which demonstrate a lack of strict periodicity resulting in a set of helical pitch lengths corresponding to the shaded region in Fig. 4(e). This is supported by recent SANS measurements of isostructural $Cr_{1/3}NbS_2$ where small site disorder results in higher order peaks. We conclude that the Mn site disorder [21] contributes significantly to the width of scattering in Fig. 4 [39].

The importance of the DMI in this system is made clear by the presence of spin textures that are both non-periodic and thickness dependent. Interestingly, magnetic contrast in Fresnel images only appears at $T < 25$ K $< T_c$=45 K, which corresponds well with changes in the *ac* susceptibility, *M(H)*, and SANS, suggesting that spin textures are intrinsic to the material, even in bulk crystalline form. The spatial extent of the spin textures in LTEM, confirmed by SANS data, is hundreds of nanometers indicating a small DMI, much smaller than was observed in isostructural $Cr_{1/3}NbS_2$ in agreement with electronic structure calculations [40]. The implication of these observations is that the DMI and the magnetostatic interactions are of similar magnitude so that variations within our samples, such as the sample thickness and crystalline disorder, although thought to be small, create significant variations in the periodicity of the spin textures. For the thin lamella explored through Fresnel images, the shape anisotropy appears to be sufficient to distort the magnetic structure, such that it is no longer helical, as the images suggest that regions where the magnetic moments lie within the plane of the lamella increase significantly for our thinnest samples.

Our experiments, together with theory and simulation, strongly support the idea that the effective sample geometry-induced anisotropy in chiral magnets can switch the magnetic states from topological to trivial, opening up a new avenue for controlling topological properties in these systems. In addition, the heterogeneity in the period of the chiral magnetic state implied by the SANS and LTEM measurements suggests that even minimal imperfections in our crystals strongly affect the local periodicity of the magnetic structure. Investigations of the crystalline structure of our samples, including x-ray and neutron diffraction and high resolution TEM imaging, do not indicate significant disorder. Far from being a deficiency, the variation we observe, instead, suggests a method to manipulate the length scales of the magnetic textures and presumably the currents necessary to drive them. Despite the lack of strict periodicity of the magnetic structure, we have demonstrated that the shape anisotropy, which tends to confine magnetic moments in the plane of thin samples, can be used to control the overall chirality of the magnetic state. Thick samples (small influence of the sample surfaces) retain the helicity defined by the DMI, and thus the topological protection of any soliton-like features, whereas thin samples do not, thereby removing this protection. The result is a demonstrated ability to influence not only the direction of motion of chiral DW, but more importantly the ability to annihilate DW of opposite chirality with magnetic field, offering an unprecedented control topological features and soliton pair dynamics.

**Acknowledgements**

The experimental material resented here is supported by the U.S. Department of Energy under EPSCoR Grant No. DE-SC0012432 with additional support from the Louisiana Board of Regents. A.K. acknowledges support for the Fresnel imaging in LTEM from the EU's ERC Horizon 2020 program under grant agreement 856538 and from DFG project-ID 405553726–TRR 270. The SANS measurements at ORNL's HFIR was sponsored by the Scientific User Facilities Division, Office of Science, Basic Energy Sciences (BES), U.S. Department of Energy (DOE). I. V. acknowledges support from NSF Grant DMR 1410741 for theoretical work, and hospitality of the KITP, where part of this research was performed under NSF Grant No. PHY-1748958. We would like to thank Dr. Ron Kelley from ThermoFisher Scientific for providing access and assistance with their PFIB for sample preparation and TEM imaging. We thank Dr. Dongmei Cao of the Shared Instrument Facilities at Louisiana State University for assistance with sample preparation for the LTEM imaging.

**Author contributions:** S.K.K, M.C., D.P.Y, and J.F.D. conceived and designed the experiments. S.K.K. synthesized crystals under the direction of D.P.Y. and performed the magnetization and *ac* magnetic susceptibility measurements. M.M. and W.X. performed the single crystal x-ray diffraction experiment. S.K.K, J.F.D., and L.D-S. performed the small-angle neutron scattering measurements, A.K. and M.C. performed the Lorentz Transmission Microscopy (LTEM) measurements. The data analysis of LTEM was performed by S.K.K with the supervision of J.F.D., M.C. and A.K. M.C. performed the micromagnetic calculations. I.V. and W.A. performed the analytic calculations. All authors participated in the writing of the manuscript.

**Emails:** skkarna@nsu.edu ; jfditusa@iu.edu

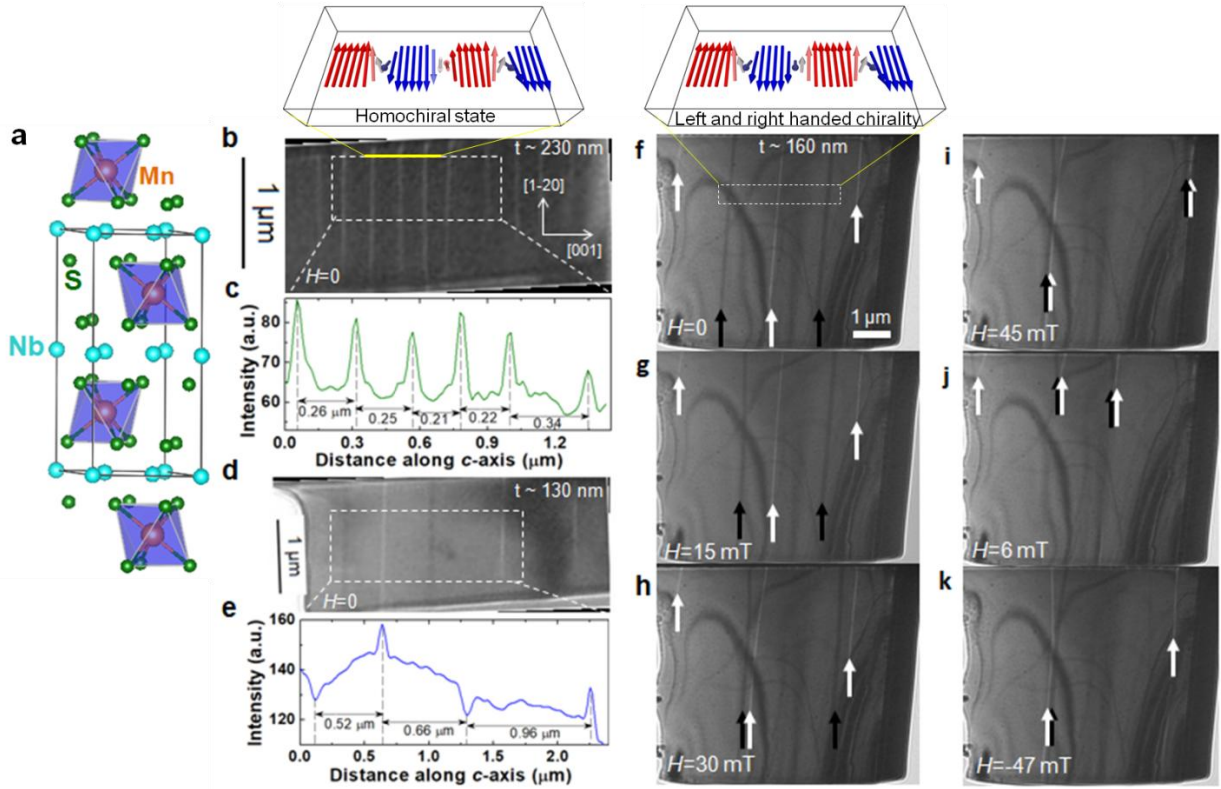

**Fig. 1** Crystal structure and magnetic microstructure of $Mn_{1/3}NbS_2$. (a) Crystal structure - the intercalated Mn atoms occupy the octahedral interstitial holes (*2c* site) between trigonal prismatic layers of 2H-$NbS_2$ in the ideal case. (b) Defocused Fresnel images for a ~230-nm thick region of sample 1 recorded at 12 K. A series of alternating bright lines (domain walls) separated by grey regions that are not strictly periodic is observed. (c) Line profile of the intensity shown in (b) integrated along the (1-20) direction for the white-boxed region shown in (b). (d) Fresnel image for a ~130-nm thick region of sample 1 recorded at 12 K. A series of alternating bright and dark lines perpendicular to the *c*-axis of the crystal are observed that lack a strict periodicity. (e) Line profile of the intensity shown in (d) integrated along the (1-20) direction for the white-boxed region shown in (b). (f)-(k) Fresnel images of sample 2 of thickness ~160 nm in zero and applied magnetic fields (identified in the figure) recorded at 14 K. Arrows indicate the position of alternating bright (white arrows) and dark (black arrows) lines of contrast. Schematics of the magnetic structure at the top of frames (b) and (f) are suggested by our micromagnetic simulations.

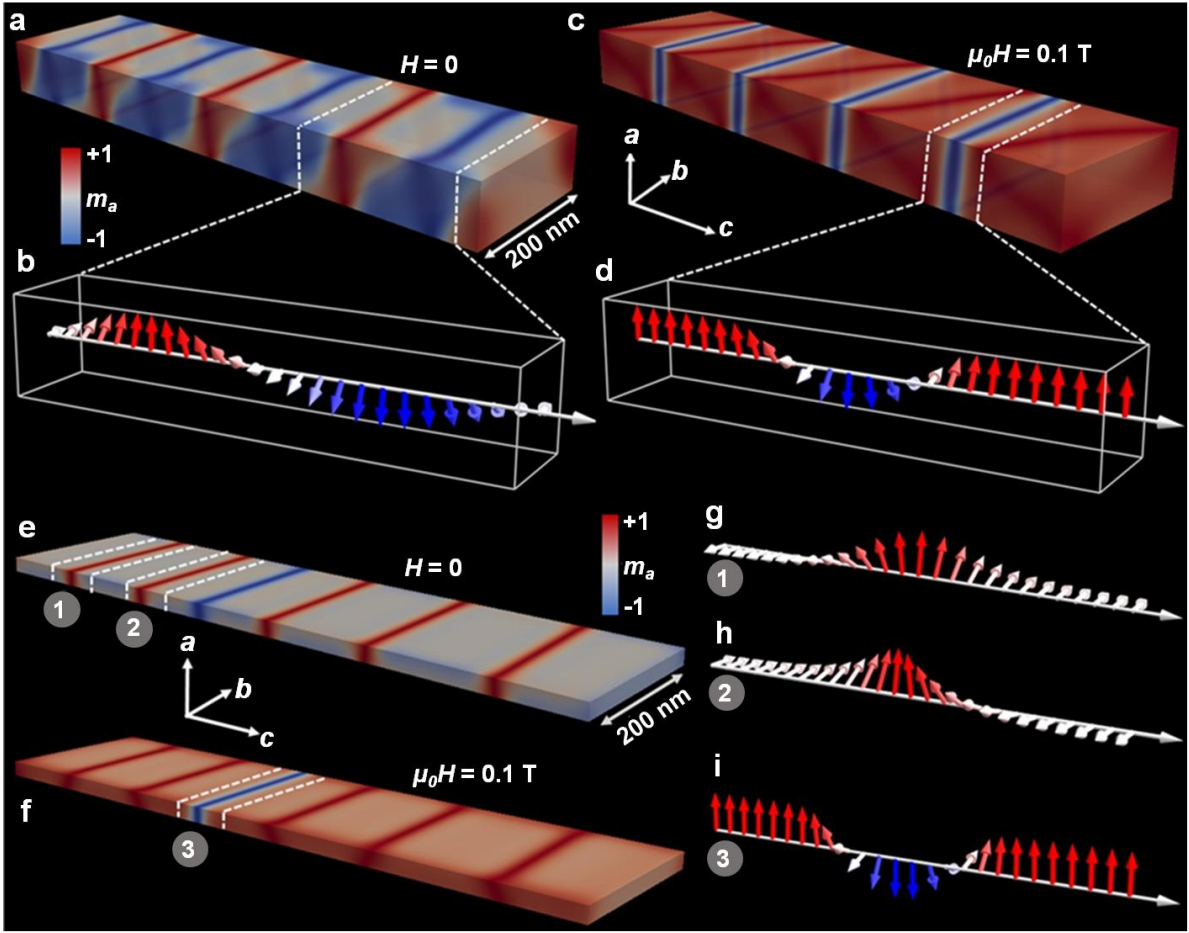

**Fig. 2** Linear soliton lattice. Snapshots from the micromagnetic simulations showing (a) a contour plot of the *a*-component of the magnetization at zero external field for a thick sample exhibiting a linear soliton lattice in the form of repeating domain-wall pairs. (b) Schematic demonstrating the generalized parameterization of the magnetization vector for the contour shown in (a). Here, each domain wall in the pair has opposite polarity, but they all have the same handedness, which is determined by the sign of the DMI. (c) Contour plot of the *a*-component of the magnetization for a thick sample in a field, $H = 0.1$ T, exhibiting a magnetic soliton lattice state. (d) Schematic highlighting a $2\pi$ right-handed chiral domain wall. If the sample is thinner, however, magnetostatic interactions play a dominant role and domain-wall pairs with the same polarity and opposing handedness occur, as shown in (e). (f) With the application of a magnetic field, a thin sample lacks much of the topological protection enjoyed by the thicker sample due to the proximity of chiral domain walls of opposite handedness. Generalized parametrization of the magnetization vector for a (g) right-handed (region 1) and (h) left-handed (region 2) π domain wall. (i) Schematic demonstration of a pair of homochiral domain wall pairs (region 3).

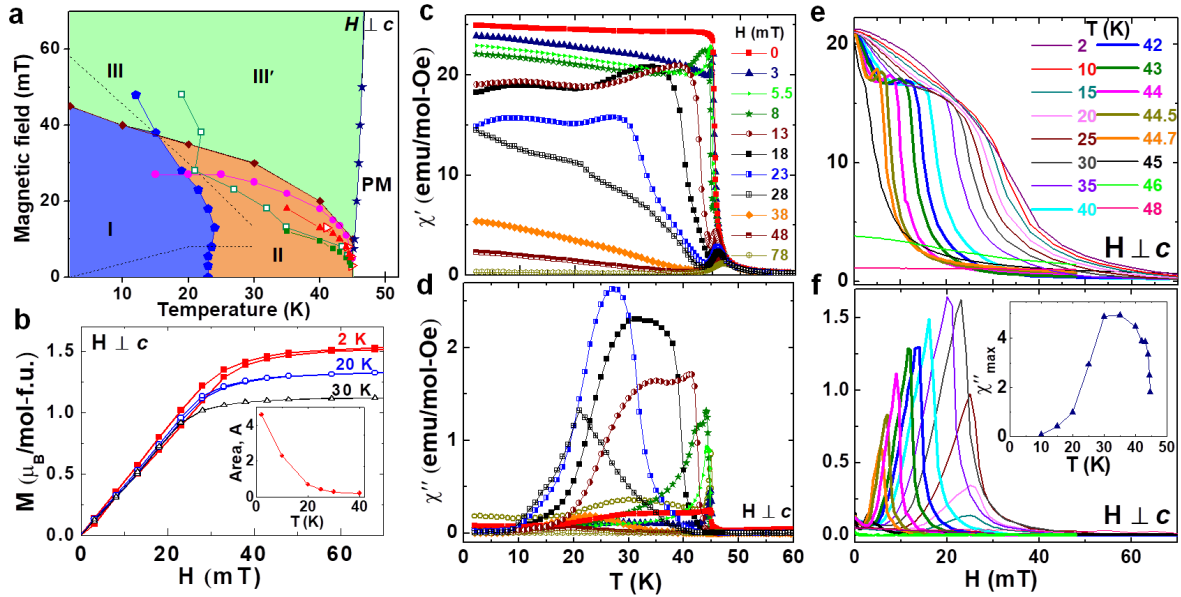

**Fig. 3** Magnetic properties. (a) Proposed magnetic phase diagram of $Mn_{1/3}NbS_2$ as a function of temperature, $T$, and magnetic field, $H$, applied perpendicular to the crystallographic $c$-axis. Phase I is a helical magnetic phase lacking strict periodicity, phase II is a nearly ferromagnetic phase that is not fully characterized, while phases III and III' are nearly fully-polarized magnetic phases, and the region above 45 K (labeled PM) is paramagnetic. The region near the critical temperature, $T_C = 45$ K, is characterized by a peak in the $T$ dependence of the real part of the ac susceptibility, $\chi'(T)$, at finite $H$ (frame (c) and Fig. S7(b) (denoted as $T_1$) and plotted in the phase diagram as solid blue stars. At lower $T$, we observe two distinct maxima in the imaginary part of the ac susceptibility, $\chi''(T)$ (frame (d) and Fig. S7(d)) that are designated in frame (a) as open red triangles ($T_2$) and open green squares ($T_3$). $T_2$ and $T_3$ are well correlated with features in the $H$ dependence of the real part of $\chi'$ (frame (e) and Fig. S7(a)), where $H_1$ (solid green squares in (a)) denotes the low field minimum and $H_2$ (solid red triangles in (a)) the maximum at slightly higher $H$. Solid pink circles indicate $H_3$, the maxima in $\chi''(H)$ (frame (f) and Fig. S7(a)), which appear

to track $T_2$ and $H_2$ at slightly higher $H$. $H_3$ also tracks the saturation field in the magnetization, $M(H)$, (frame (b) and designated by purple diamonds in frame (a)) but at a somewhat smaller $H$. At lower temperatures there is a hysteresis observed in $M(H)$ (frame (b)) with a range indicated by the dotted lines in frame (a). The upper temperature limit of the hysteretic region correlates well over a range of the phase diagram with the maximum in the derivative of $\chi''(T)$ with respect to $T$ ($d\chi''/dT$) (frame (f) and the inset to Fig. S7(c)), which is indicted in frame (a) as solid blue pentagons ($T_4$). Inset to frame (f): $T$ dependence of the magnitude of the maximum in $\chi''(H)$ shown in (f).

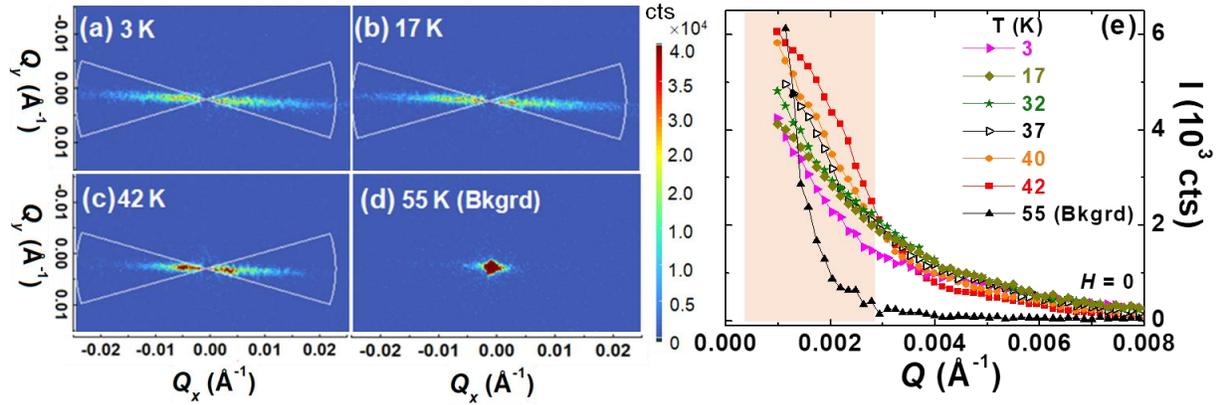

**Fig. 4** Small Angle Neutron Scattering (SANS) measurement of $Mn_{1/3}NbS_2$ at (a) 3 K, (b) 17 K, and (c) 42 K with the wavevector, $Q$, parallel to the (001) reflection along the horizontal. The signal at 55 K (frame (d)) was considered as background and subtracted from the data collected at lower temperatures. Data were obtained with the incident neutron beam perpendicular to the *c*-axis. (e) Variation of the integrated intensity, $I$, obtained from the area designated by the white lines in frames (a)-(c) *vs.* $Q$ at the indicated temperatures. The shaded region corresponds to the $Q$-range expected for the periodicities found in the Lorentz TEM studies (see Fig 1).